# Which are the influential publications in the Web of Science subject categories over a long period of time? CRExplorer software used for big-data analyses in bibliometrics

Andreas Thor[1], Lutz Bornmann[2], Robin Haunschild[3], Loet Leydesdorff[4]

[1] *thor@hft-leipzig.de*
University of Applied Sciences for Telecommunications Leipzig, Gustav-Freytag-Str. 43-45, 04277 Leipzig (Germany).

[2] *bornmann@gv.mpg.de*
Division for Science and Innovation Studies, Administrative Headquarters of the Max Planck Society, Hofgartenstr. 8, 80539 Munich (Germany).

[3] *r.haunschild@fkf.mpg.de*
Max Planck Institute for Solid State Research, Information Service, Heisenbergstrasse 1, 70506 Stuttgart, (Germany).

[4] *loet@leydesdorff.net*
University of Amsterdam, P.O. Box 15793, 1001 NG Amsterdam (The Netherlands).

**Abstract**
What are the landmark papers in scientific disciplines? On whose shoulders does research in these fields stand? Which papers are indispensable for scientific progress? These are typical questions which are not only of interest for researchers (who frequently know the answers – or guess to know them), but also for the interested general public. Citation counts can be used to identify very useful papers, since they reflect the wisdom of the crowd; in this case, the scientists using the published results for their own research. In this study, we identified with recently developed methods for the program CRExplorer landmark publications in nearly all Web of Science subject categories (WoSSCs). These are publications which belong more frequently than other publications across the citing years to the top-‰ in their subject category. The results for three subject categories "Information Science & Library Science", "Computer Science, Information Systems", and "Computer Science, Software Engineering" are exemplarily discussed in more detail. The results for the other WoSSCs can be found online at http://crexplorer.net.

**Introduction**

Bibliometrics is frequently used in research evaluation. In an overview, Sivertsen (2017) notes that bibliometric indicators are considered in many national research-funding systems in the European Union to measure research performance. Not only researchers themselves, but also science administrators and the public are interested in reports on groundbreaking research from units of assessments (e.g., universities or countries) (e.g., van Noorden, Maher, & Nuzzo, 2014). According to Winnink, Tijssen, and van Raan (2018), the term groundbreaking (or breakthrough) is often used for research (discoveries) with a major impact on future scientific activities. Hollingsworth (2008) considers breakthroughs as very useful to many researchers in targeting future research questions in various scientific fields.

Although breakthroughs are of general interest in science (Orduna-Malea, Martín-Martín, & Delgado López-Cózar, 2018; Schlagberger, Bornmann, & Bauer, 2016), research evaluation focuses – as a rule – on short-time horizons: "the time horizon is 10 years or less, and the focus is on recent past performance, as it is believed to increase the policy relevance, and reduce data collection costs" (Moed, 2017, p. 6). Whereas short-term impact measurements allow statements about the research front, "long-term impact indicates to what extent they eventually succeed in scoring 'triumphs'" (Moed, Burger, Frankfort, & van Raan, 1985, p. 134). The results of Wang (2013) further show that the frequent use of a short citation window (the standard is a minimum of three years) may lead to hasty classifications of papers as high-impact papers which turn out to be wrong in the long run (Baumgartner & Leydesdorff, 2014;

Ponomarev, Williams, Hackett, Schnell, & Haak, 2014). The results of Wang, Veugelers, and Stephan (2017) and Mairesse and Pezzoni (2018) reveal that novel papers are associated with high citation rates especially in the long run.

Winnink et al. (2018) studied five algorithms for detecting breakthrough papers. The results point out that the algorithms are powerful tools for tracing breakthrough papers. van Noorden et al. (2014) used traditional citation analyses to identify the most cited publications of all time. They found that about 15,000 papers have more than 1,000 citations and thus seem to be very useful. Marx, Bornmann, Barth, and Leydesdorff (2014) developed the method Reference Publication Year Spectroscopy (RPYS) to detect the origins of research fields or topics. The method is based on counting cited references (instead of citations) to assess the impact of publications on a topic- or field-specific publication set (e.g., climate change, see Marx, Haunschild, Thor, & Bornmann, 2017). The method has already been successfully applied in identifying papers with outstanding performance (Comins & Leydesdorff, 2018; Thor, Bornmann, Marx, & Mutz, 2018) and landmark patents (Comins, Carmack, & Leydesdorff, 2017).

Thor, Marx, Leydesdorff, and Bornmann (2016) introduced the CRExplorer – a program for undertaking RPYS. In a recent update of the program, Thor et al. (2018) developed an indicator for identifying publications in research fields which are influential over longer periods. In other words, publications (cited references) can be identified which belong to the 10% most-referenced publications in many citing years. In this study, we run the CRExplorer on a powerful computer and use a new variant of the indicator to identify publications which belong to the 1‰ (0.1%) most-referenced publications in all citing years between 1980 and 2017 in 205 subject categories (named as N_TOP0_1+). With focusing on the top-‰, we have identified the exceptionally useful shoulders on which published research in the subject categories between 1980 and 2017 stood. In this paper, the procedure is explained how the shoulders have been identified. The results for three subject categories are explained in this paper in more detail; the results for all subject categories can be inspected online at http://crexplorer.net.

## Methods

*Datasets used*

We used the Web of Science (WoS, Clarivate Analytics) custom data of the Max Planck Society's in-house database derived from the Science Citation Index Expanded (SCI-E), Social Sciences Citation Index (SSCI), and Arts and Humanities Citation Index (AHCI) produced by Clarivate Analytics (Philadelphia, USA). All records for the papers of the document type "article" published between 1980 and 2017 were exported separately for each WoS subject category (WoSSC). The WoSSCs were ordered by their number of publications from CQ ("Biochemistry & Molecular Biology") with 1,455,479 articles to 9a ("Green & Sustainable Science & Technology") with 3,169 articles (see Leydesdorff, 2006). We required a ratio of linked vs. cited references of at least 0.30 for a WoSSC to be included. The reason is that only WoSSCs with sufficient references covered by the WoS should be considered in the analyses. In total, 205 WoSSCs were considered.

*Indicator used*

We are interested in those cited references which have been (statistically) significantly cited more frequently in the citing years than other cited references in the dataset. To this end, for each cited reference we count the number of citing years where the cited reference has been cited extraordinarily frequently. For each citing year, all $n$ cited references have been sorted in descending order based on their citation counts in the citing year. We then identified the citation count $c$ of the cited reference at rank ($1+n/1000$), i.e., the cited reference that follows the first (top) 0.1% cited references. For example, for $n=10.000$ cited references we determined the number of citations of the cited reference at rank #11. All cited references with a citation count greater than $c$ are then considered as "top cited reference" in the citing year if their citation count is additionally above the average of the expected citation count (see Thor et al., 2018, for details on the sequence computation). The metric N_TOP0_1+ is the number of citing years where the cited reference is a "top cited reference".

*CRExplorer script*

The following CRExplorer script was used to perform the RPYS and filter for exceptionally highly referenced publications for each of the WoSSCs:

```
set(n_pct_range: 2, median_range: 2)
importFile(file: "xx_wos.txt", type: "WOS",
RPY: [1900, 2015, false], PY: [1980, 2017, false], maxCR: 0)
info()
cluster(threshold: 0.75, volume: true, page: true, DOI: false)
merge()
exportFile(file: "xx_wos.rpys_CR.csv", type: "CSV_CR",
     sort: ["N_TOP0_1_Plus DESC", "N_CR DESC"],
filter: { it.N_TOP0_1_Plus >= 10 } )
```

Listing 1: CRExplorer script to perform RPYS and filter for cited references with an indicator value of at least 10 for N_TOP0_1+

Two neighboring years are included in the calculation of the advanced indicators via the set options. Thus, not only the focal years are considered in the calculation, but also neighboring years to increase the case numbers for the analyses. The file name "xx_wos.txt" has to be adjusted for each WoSSC in the importFile function. The PY option ensures that only papers published between 1980 and 2017 are included. The RPY option guarantees that only cited references published between 1900 and 2015 are included. We expect no exceptionally highly referenced papers before 1900. We also expect that cited references published after 2015 did not have enough time to become exceptionally highly referenced, especially in many citing years. The clustering and merging of variants of the same cited reference in the dataset is done with the Levenshtein threshold of 0.75 including volume and page but not DOI in the cited references' information (Thor et al., 2016). The file name "xx_wos.rpys_CR.csv" in the exportFile function has to be adjusted for each WoSSC. In addition, this function filters for cited references with an indicator value of at least 10 and sorts the results according to the indicator value and the number of cited references before writing the cited references into the CSV file. The value of 10 has been adjusted to a lower one if cited references in some WoSSCs do not achieve large enough indicator values. For the WoSSCs with many papers and many cited references variants, we needed 382 GB of main memory (RAM).

**Results**

The identified landmark papers for nearly all WoSSC can be inspected online at http://crexplorer.net (see Figure 1).

| WoSSC Code | WoSSC name | CR | RPY | N_CR | N_TOP0_1+ | Link |
|---|---|---|---|---|---|---|
| nu | Information Science & Library Science | PORTER ME, 1980, COMPETITIVE STRATEGY | 1980 | 173 | 20 | |
| nu | Information Science & Library Science | GIDDENS A, 1984, CONSTITUTION SOC | 1984 | 136 | 19 | |
| nu | Information Science & Library Science | BELKIN NJ, 1982, JOURNAL OF DOCUMENTATION, V38, P61 | 1982 | 309 | 18 | |
| nu | Information Science & Library Science | VANRIJSBERGEN CJ, 1979, INFORMATION RETRIEVA | 1979 | 281 | 18 | |
| nu | Information Science & Library Science | WHITE HD, 1981, JOURNAL OF THE AMERICAN SOCIETY FOR INFORMATION SCIENCE, V32, P163 | 1981 | 223 | 18 | http://gateway.proquest.com/openurl?res_dat=xri%3Apqm&volume=32&date=1981&rft_val_f |
| nu | Information Science & Library Science | Beaver D. deB., 1978, Scientometrics, V1, P65 | 1978 | 134 | 18 | https://link.springer.com/article/10.1007/BF02016840 |
| nu | Information Science & Library Science | PORTER MF, 1980, PROGRAM-AUTOMATED LIBRARY AND INFORMATION SYSTEMS, V14, P130 | 1980 | 287 | 17 | https://www.emeraldinsight.com/doi/abs/10.1108/eb046814 |
| nu | Information Science & Library Science | MARKUS ML, 1983, COMMUNICATIONS OF THE ACM, V26, P430 | 1983 | 271 | 17 | https://oadoi.org/10.1145/358141.358148 |
| nu | Information Science & Library Science | ROGERS EM, 1983, DIFFUSION INNOVATION | 1983 | 261 | 17 | |
| nu | Information Science & Library Science | CHURCHILL GA, 1979, JOURNAL OF MARKETING RESEARCH, V16, P64 | 1979 | 247 | 17 | |
| dk | Business, Finance | WHITE H, 1980, ECONOMETRICA, V48, P817 | 1980 | 2312 | 23 | http://www.jstor.org/stable/1912934?origin=crossref |
| dk | Business, Finance | DICKEY DA, 1979, JOURNAL OF THE AMERICAN STATISTICAL ASSOCIATION, V74, P427 | 1979 | 543 | 23 | http://ezb.uni-regensburg.de/ezeit/warpto.phtml?bibid=MPIS&color=2&jour_id=7757&url=htt |
| dk | Business, Finance | HANSEN LP, 1982, ECONOMETRICA, V50, P1029 | 1982 | 852 | 22 | http://www.jstor.org/stable/1912775?origin=crossref |
| dk | Business, Finance | VASICEK O, 1977, JOURNAL OF FINANCIAL ECONOMICS, V5, P177 | 1977 | 578 | 22 | https://doi.org/10.1016/0304-405X(77)90016-2?nols=y&urlappend=%3Fgoto%3Dsd |
| dk | Business, Finance | HARRISON JM, 1979, JOURNAL OF ECONOMIC THEORY, V20, P381 | 1979 | 426 | 22 | https://doi.org/10.1016/0022-0531(79)90043-7?nols=y&urlappend=%3Fgoto%3Dsd |
| dk | Business, Finance | BLACK F, 1973, JOURNAL OF POLITICAL ECONOMY, V81, P637 | 1973 | 2288 | 21 | https://openurl.ebscohost.com/linksvc/linking.aspx?sid=buh&volume=81&aulast=BLACK&atitle |
| dk | Business, Finance | LUCAS RE, 1978, ECONOMETRICA, V46, P1429 | 1978 | 455 | 21 | http://www.jstor.org/stable/1913837?origin=crossref |
| dk | Business, Finance | DICKEY DA, 1981, ECONOMETRICA, V49, P1057 | 1981 | 338 | 19 | http://www.jstor.org/stable/1912517?origin=crossref |
| dk | Business, Finance | MERTON RC, 1973, BELL JOURNAL OF ECONOMICS, V4, P141 | 1973 | 1067 | 18 | https://oadoi.org/10.2307/3003143 |
| dk | Business, Finance | BLACK F, 1976, JOURNAL OF FINANCIAL ECONOMICS, V3, P167 | 1976 | 423 | 18 | https://doi.org/10.1016/0304-405X(76)90024-6?nols=y&urlappend=%3Fgoto%3Dsd |
| my | Psychology, Developmental | REYNOLDS CR, 1978, JOURNAL OF ABNORMAL CHILD PSYCHOLOGY, V6, P271 | 1978 | 431 | 23 | http://gateway.proquest.com/openurl?res_dat=xri%3Apqm&volume=6&date=1978&rft_val_fm |
| my | Psychology, Developmental | SPANIER GB, 1976, JOURNAL OF MARRIAGE AND THE FAMILY, V38, P15 | 1976 | 380 | 23 | https://openurl.ebscohost.com/linksvc/linking.aspx?sid=sih&volume=38&aulast=SPANIER&atitl |
| my | Psychology, Developmental | BECK AT, 1961, ARCHIVES OF GENERAL PSYCHIATRY, V4, P561 | 1961 | 569 | 22 | https://1findr.1science.com/search?query=10.1001/archpsyc.1961.01710120031004 |
| my | Psychology, Developmental | ROSENBERG M, 1965, SOC ADOLESCENT SELF | 1965 | 945 | 21 | |
| my | Psychology, Developmental | COHEN J, 1960, EDUCATIONAL AND PSYCHOLOGICAL MEASUREMENT, V20, P37 | 1960 | 692 | 21 | http://journals.sagepub.com/doi/abs/10.1177/001316446002000104 |
| my | Psychology, Developmental | BOWLBY J, 1973, ATTACHMENT LOSS, V2 | 1973 | 675 | 20 | |
| my | Psychology, Developmental | BRONFENBRENNER U, 1979, ECOLOGY HUMAN DEV | 1979 | 596 | 20 | |
| my | Psychology, Developmental | STRAUS MA, 1979, JOURNAL OF MARRIAGE AND THE FAMILY, V41, P75 | 1979 | 338 | 20 | https://openurl.ebscohost.com/linksvc/linking.aspx?sid=sih&volume=41&aulast=STRAUS&atitle |
| my | Psychology, Developmental | HARTER S, 1982, CHILD DEVELOPMENT, V53, P87 | 1982 | 608 | 19 | |
| my | Psychology, Developmental | KOVACS M, 1981, ACTA PAEDOPSYCHIATRICA, V46, P305 | 1981 | 441 | 19 | |
| lq | Health Policy & Services | ANDERSEN R, 1973, MILBANK MEMORIAL FUND QUARTERLY-HEALTH AND SOCIETY, V51, P95 | 1973 | 320 | 20 | http://www.jstor.org/stable/3349613?origin=crossref |
| lq | Health Policy & Services | FOLSTEIN MF, 1975, JOURNAL OF PSYCHIATRIC RESEARCH, V12, P189 | 1975 | 423 | 19 | https://doi.org/10.1016/0022-3956(75)90026-6?nols=y&urlappend=%3Fgoto%3Dsd |
| lq | Health Policy & Services | NUNNALLY JC, 1978, PSYCHOMETRIC THEORY | 1978 | 246 | 19 | |
| lq | Health Policy & Services | WHITE H, 1980, ECONOMETRICA, V48, P817 | 1980 | 253 | 18 | http://www.jstor.org/stable/1912934?origin=crossref |
| lq | Health Policy & Services | HECKMAN JJ, 1979, ECONOMETRICA, V47, P153 | 1979 | 194 | 17 | http://www.jstor.org/stable/1912352?origin=crossref |
| lq | Health Policy & Services | RAWLS J, 1971, THEORY JUSTICE | 1971 | 147 | 17 | |
| lq | Health Policy & Services | RADLOFF L S, 1977, Applied Psychological Measurement, V1, P385 | 1977 | 644 | 16 | http://journals.sagepub.com/doi/abs/10.1177/014662167700100306 |
| lq | Health Policy & Services | Aday L A, 1974, Health services research, V9, P208 | 1974 | 200 | 16 | |
| lq | Health Policy & Services | SHROUT PE, 1979, PSYCHOLOGICAL BULLETIN, V86, P420 | 1979 | 196 | 16 | http://doi.apa.org/getdoi.cfm?doi=10.1037%2F0033-2909.86.2.420 |
| lq | Health Policy & Services | ENDICOTT J, 1976, ARCHIVES OF GENERAL PSYCHIATRY, V33, P766 | 1976 | 158 | 16 | |
| wh | Rheumatology | TAN EM, 1982, ARTHRITIS AND RHEUMATISM, V25, P1271 | 1982 | 4849 | 20 | https://1findr.1science.com/search?query=10.1002/art.1780251101 |

**Figure 1. Online presentation of the landmark papers**

In the following we focus exemplarily on three WoSSCs and explain the results in more detail. We selected WoSSCs which we are able to interpret based on our own field-specific expertises. Table 1 shows the results for the WoSSC "Information Science & Library Science". Five cited publications are listed exemplarily with the most citing years in which the publication belongs to the top-‰. Two publications in the table are basic works on information retrieval (Belkin, Oddy, & Brooks, 1982; Van Rijsbergen, 1979). Three of the five publications in the table are not primarily contributions to the library and information science (LIS) field: Michael Porter's (1980) book is one of his contributions to the field of business economics. In later work, Porter (1990) became specifically known for cluster analysis in the follow-up book entitled "The Competitive Advantage of Nations." Anthony Giddens' (1984) book entitled "The Constitution of Society" is the *locus classicus* of Giddens' "structuration theory" in sociology. Both this book and Porter (1980) are well known and intensively used outside the specialist's communication. Both books are theoretical, but oriented towards application (without providing a methodology). White and Griffith (1981) introduced author-co-citation analysis (ACA) in LIS and Science & Technology Studies. ACA became thereafter a widely used technique. It is primarily a statistical method, but it can also be used in a qualitative analysis.

**Table 1. Most exceptionally referenced cited references
in the WoSSC "Information Science & Library Science".**

| RPY | CR | N_CR | N_TOP0_1+ |
|---|---|---|---|
| 1980 | Porter, M. E.: *Competitive Strategy: Techniques for Analyzing Industries and Competitors*. Free Press | 173 | 20 |
| 1984 | Giddens, A.: *The Constitution* of *Society*. Outline of the Theory of Structuration. Polity Press | 136 | 19 |
| 1982 | *Belkin*, *N. J.*, Oddy, R. N. and Brooks, H.: ASK for Information Retrieval: Part I. Background and Theory. *Journal of Documentation*, *38*(2), *61-71* | 309 | 18 |

| 1979 | Van Rijsbergen, C.J.: *Information Retrieval*. Unpublished PhD thesis, Department of Computing Science, University of Glasgow | 281 | 18 |
| 1981 | White, H. D., & Griffith, B. C.: Author Cocitation - a Literature Measure of Intellectual Structure. *Journal of the American Society for Information Science, 32*(3), 163-171 | 223 | 18 |

Notes. RPY=Reference publication year; CR=Cited reference; N_CR=Number of cited references; N_TOP0_1+=Number of citing years in which the publication belongs to the top-‰.

Table 2 shows the results for the WoSSC "Computer Science, Information Systems". The three papers "A Method for obtaining digital Signatures and public-key Cryptosystems" (Rivest, Shamir, & Adleman, 1978), "A public-key Cryptosystem and a Signature Scheme based on discrete Logarithms" (ElGamal, 1985), and "New Directions in Cryptography" (Diffie & Hellman, 2006) describe fundamental algorithms for data encryption and digital signatures. These algorithms are important for secure (i.e., encrypted) data transmission over the Internet. The idea of an asymmetric cryptosystem based on public and private keys (that can be exchanged securely) is used in current software such as PGP. Rivest et al. (1978) also received the ACM Turing award (the "Nobel prize for computer science") for their work. The book by Garey and Johnson (1979) "Computers and Intractability: a Guide to the Theory of NP-Completeness" gives an introduction to computational complexity, a fundamental concept in theoretical computer science. The book is well-known for its extensive list of NP-complete problems, i.e., problems where an efficient solution (i.e., in polynomial time) does not yet exist. Especially in the era of big data, efficient software algorithms (besides large clusters of hardware components) are a cornerstone of many web applications. "The Theory of error-correcting Codes" (MacWilliams & Sloane, 1977) is an influencing book on information theory and coding theory. It describes approaches for the reliable transmission of data over unreliable communication channels, e.g., when multiple mobile phones interfere with each other on the same WiFi network.

Table 2. Most exceptionally referenced cited references
in the WoSSC "Computer Science, Information Systems".

| RPY | CR | N_CR | N_TOP0_1+ |
|---|---|---|---|
| 1978 | Rivest, R. L., Shamir, A., & Adleman, L. (1978). A Method for obtaining digital Signatures and public-key Cryptosystems. *J Commun. ACM, 21*(2), 120-126 | 862 | 21 |
| 1979 | Garey, M. R. & Johnson, D. S. (1979). *Computers and Intractability: A Guide to the Theory of NP-completeness*. W. H. Freeman | 1137 | 19 |
| 1977 | MacWilliams, F. J., & Sloane, N. J. A. (1977). *The Theory of error Correcting Codes*. North-Holland Publishing Company | 689 | 19 |
| 1985 | ElGamal, T. (1985). *A public key Cryptosystem and a Signature Scheme Based on discrete Logarithms*. Paper presented at the Workshop on the Theory and Application of Cryptographic Techniques, Berlin, Heidelberg | 503 | 19 |

| RPY | CR | N_CR | N_TOP0_1+ |
|---|---|---|---|
| 1976 | Diffie, W., & Hellman, M. (2006). New Directions in Cryptography *J IEEE Trans. Inf. Theor, 22*(6), 644-654 | 878 | 18 |

Notes. RPY=Reference publication year; CR=Cited reference; N_CR=Number of cited references; N_TOP0_1+=Number of citing years in which the publication belongs to the top-‰.

The results for the WoSSC "Computer Science, Software Engineering" are reported in Table 3. The first two cited references are the in area of theoretical computer science. The book by Garey and Johnson (1979) has already been described since it also appears in the top list of "Computer Science, Information Systems". The paper "Maintaining Knowledge about temporal Intervals" (Allen, 1983) introduces a calculus for temporal reasoning. This is important for software or robots using artificial intelligence where the concept of time (i.e., when things happen) is important. The two papers "Recursively generated B-spline Surfaces on arbitrary topological Meshes" (Catmull & Clark, 1978) and "Theory of Edge Detection" (Marr, Hildreth, & Brenner, 1980) are in the area of computer graphics. The technique of B-spline surfaces is used in computer graphics to create smooth surfaces. This is, for example, important in 3D video games to generate realistically looking objects. Edge detection is a core task in processing digital images to detect and extract features (e.g., objects) in digital images. This is particularly important in computed tomography technique (CT) to detect objects of interest, e.g., arteries. Weiser (1984) introduced the concept of "Program slicing", a method for automatically decomposing programs into so-called slices. The decomposition can be used for efficient finding of errors (debugging) but also for software maintenance and optimization. Though the concept has been significantly extended over the years, it is still a fundamental concept in professional software engineering.

**Table 3. Most exceptionally referenced cited references in the WoSSC "Computer Science, Software Engineering".**

| RPY | CR | N_CR | N_TOP0_1+ |
|---|---|---|---|
| 1979 | Garey, M. R. & Johnson, D. S. (1979). *Computers and Intractability: A Guide to the Theory of NP-completeness.* W. H. Freeman | 867 | 19 |
| 1983 | Allen, J. F. (1983). Maintaining knowledge about temporal Intervals. *J Commun. ACM, 26*(11), 832-843 | 231 | 19 |
| 1978 | Catmull, E., & Clark, J. (1978). Recursively generated B-spline Surfaces on arbitrary topological Meshes. *Computer-Aided Design, 10*(6), 350-355 | 364 | 18 |
| 1980 | Marr, D., Hildreth, E., & Brenner, S. (1980). Theory of Edge Detection. *207*(1167), 187-217 | 206 | 18 |
| 1984 | Weiser, M. (1984). Program slicing. *IEEE Transactions on Software Engineering, SE-10*(4), 439-449 | 351 | 17 |

Notes. RPY=Reference publication year; CR=Cited reference; N_CR=Number of cited references; N_TOP0_1+=Number of citing years in which the publication belongs to the top-‰.

**Discussion**

What are the landmark papers in scientific fields? On whose shoulders does research in these fields stand? Which papers would be indispensable for scientific progress? These are typical questions which are not only of interest for researchers (who frequently know the answers – or are supposed to know them), but also for the general public (e.g., science journalists). Citation counts are often used to identify very useful papers, since they reflect the wisdom of the crowd; in this case, the many scientists citing the published results in their own papers. The problem with today's research evaluation processes is, however, that they focus on rather recent years (the last few years) to assess the recent developments. This focus might be able to identify research at the research front which is short-term oriented, but neglect research which appears successful in the long run. Extreme representatives of delayed recognition are so-called "sleeping beauties" which are not or scarcely cited during many years, but are heavily cited after a decade or so. These papers become useful only many years after the research has been finished.

In this study, we identified landmark publications in 205 WoSSCs with recently developed methods for the program CRExplorer. These are publications which belong more frequently than other publications to the top-‰ in their subject category across the citing years. In this paper, the results for the three WoSSCs "Information Science & Library Science", "Computer Science, Information Systems", and "Computer Science, Software Engineering" have been discussed in more detail. The results for nearly all WoSSCs can be found online (see http://crexplorer.net). It was only possible with a very powerful computer to generate the results for very large WoSSCs in our dataset. Since most users of the CRExplorer do not have these computers for undertaking cited references analyses, we deem it useful for researchers in various fields, science administrators, science journalists, and other people from the general public to have access to these landmark papers' lists.

The identification of very useful research based on citations (or cited references) is based on the premise that citations measure usefulness. Recent research suggests that citations reflect "appropriateness" which supports the use of citations in science studies and evaluation practices (Wang, 2014). However, citations are not able to reflect all influences which were useful for extraordinary research (the later landmark papers). It is especially relevant for extraordinary research to be influenced by many channels to receive this specific status: "Take Darwin. Many scholars have emphasized that although Darwin was a recluse, he was not only a voracious reader of the scientific literature but maintained a massive worldwide correspondence with explorers, naturalists, and researchers (Burkhardt, 1985–2014). Among this correspondence, Darwin received a manuscript from the Malay Archipelago entitled 'On the tendency of varieties to depart indefinitely from the original type,' which finally prodded him into publishing On the Origin of Species the following year" (MacRoberts & MacRoberts, 2017, pp. 474-475). Another problem is the incompleteness of many reference lists: "No one who has read J. D. Watson's (1968) personal account of the discovery of the structure of DNA can ever accept that the six references listed at the end of the famous Watson and Crick 1953 paper in Nature reflect the influence on their discovery … It is also clear from all accounts that, by 1952, it was the informal level of communication that was important. It was what the scientists were doing on the moving edge of research/speculation that was important to Watson and Crick, and they made every effort to get that information. Clearly, the Watson and Crick paper, similar to all scientific papers, is a 'misrepresentation' of what scientists actually do" (MacRoberts & MacRoberts, 2017, p. 475).

Against the backdrop of the critique of using citations in research evaluation purposes, the generated lists of landmark publications should only be used as hints to possible landmark publications. Users of the lists should be experts in the fields (or should consult experts) who can compare the results with their own beliefs of landmark papers. For example, in the

"Information Science & Library Science" field, the results seem counter-intuitive (against the backdrop of our expert knowledge). One would not expect Porter (1980) and Giddens (1984) to head the ranks. However, one should consider in the interpretation of the results presented in this paper and online at http://crexplorer.net that only up to ten classic papers are presented and many others follow which are (somewhat) lower ranked.

**Acknowledgments**

**References**

Allen, J. F. (1983). Maintaining knowledge about temporal intervals. *J Commun. ACM, 26*(11), 832-843. doi: 10.1145/182.358434.
Baumgartner, S. E., & Leydesdorff, L. (2014). Group-based trajectory modeling (GBTM) of citations in scholarly literature: Dynamic qualities of "transient" and "sticky knowledge claims". *Journal of the Association for Information Science and Technology, 65*(4), 797-811. doi: 10.1002/asi.23009.
Belkin, N. J., Oddy, R. N., & Brooks, H. M. (1982). Ask for Information-Retrieval .1. Background and Theory. *Journal of Documentation, 38*(2), 61-71. doi: DOI 10.1108/eb026722.
Catmull, E., & Clark, J. (1978). Recursively generated B-spline surfaces on arbitrary topological meshes. *Computer-Aided Design, 10*(6), 350-355. doi: https://doi.org/10.1016/0010-4485(78)90110-0.
Comins, J., & Leydesdorff, L. (2018). Data-mining the Foundational Patents of Photovoltaic Materials: An application of Patent Citation Spectroscopy. Retrieved April 27, 2018, from https://arxiv.org/abs/1801.09479
Comins, J. A., Carmack, S. A., & Leydesdorff, L. (2017). Patent Citation Spectroscopy (PCS): Algorithmic retrieval of landmark patents. Retrieved November 15, 2017, from https://arxiv.org/abs/1710.03349
Diffie, W., & Hellman, M. (2006). New directions in cryptography. *J IEEE Trans. Inf. Theor, 22*(6), 644-654. doi: 10.1109/tit.1976.1055638.
ElGamal, T. (1985). *A Public Key Cryptosystem and a Signature Scheme Based on Discrete Logarithms.* Paper presented at the Workshop on the Theory and Application of Cryptographic Techniques, Berlin, Heidelberg.
Garey, M. R., & Johnson, D. S. (1979). *Computers and Intractability: A Guide to the Theory of NP-completeness*: W. H. Freeman.
Giddens, A. (1984). *The Constitution of Society: Outline of the Theory of Structuration*. Cambridge: Polity Press.
Hollingsworth, J. R. (2008). Scientific discoveries: an institutionalist and path-dependent perspective. In C. Hannaway (Ed.), *Biomedicine in the Twentieth Century: Practices, Policies, and Politics, volume 72 of Biomedical and Health Research* (pp. 317–353). Bethesda, MD, USA: National Institutes of Health.
Leydesdorff, T. (2006). Can scientific journals be classified in terms of aggregated journal-journal citation relations using the Journal Citation Reports? *Journal of the American Society for Information Science and Technology, 57*(5), 601-613.
MacRoberts, M. H., & MacRoberts, B. R. (2017). The mismeasure of science: Citation analysis. *Journal of the Association for Information Science and Technology, 69*(3), 474-482. doi: 10.1002/asi.23970.
MacWilliams, F. J., & Sloane, N. J. A. (1977). *The Theory of Error Correcting Codes*: North-Holland Publishing Company.

Mairesse, J., & Pezzoni, M. (2018). Novelty in Science: The impact of French physicists' novel articles. In P. Wouters (Ed.), *Proceedings of the science and technology indicators conference 2018 Leiden "Science, Technology and Innovation indicators in transition"*. Leider, the Netherlands: University of Leiden.

Marr, D., Hildreth, E., & Brenner, S. (1980). Theory of edge detection. *207*(1167), 187-217. doi: doi:10.1098/rspb.1980.0020.

Marx, W., Bornmann, L., Barth, A., & Leydesdorff, L. (2014). Detecting the historical roots of research fields by reference publication year spectroscopy (RPYS). *Journal of the Association for Information Science and Technology, 65*(4), 751-764. doi: 10.1002/asi.23089.

Marx, W., Haunschild, R., Thor, A., & Bornmann, L. (2017). Which early works are cited most frequently in climate change research literature? A bibliometric approach based on Reference Publication Year Spectroscopy. *Scientometrics, 110*(1), 335-353. doi: 10.1007/s11192-016-2177-x.

Moed, H. F. (2017). *Applied Evaluative Informetrics*. Heidelberg, Germany: Springer.

Moed, H. F., Burger, W. J. M., Frankfort, J. G., & van Raan, A. F. J. (1985). The use of bibliometric data for the measurement of university research performance. *Research Policy, 14*(3), 131-149.

Orduna-Malea, E., Martín-Martín, A., & Delgado López-Cózar, E. (2018). Classic papers: using Google Scholar to detect the highly-cited documents. In P. Wouters (Ed.), *Proceedings of the science and technology indicators conference 2018 Leiden "Science, Technology and Innovation indicators in transition"*. Leider, the Netherlands: University of Leiden.

Ponomarev, I. V., Williams, D. E., Hackett, C. J., Schnell, J. D., & Haak, L. L. (2014). Predicting highly cited papers: A Method for Early Detection of Candidate Breakthroughs. *Technological Forecasting and Social Change, 81*(0), 49-55. doi: 10.1016/j.techfore.2012.09.017.

Porter, M. E. (1980). *Competitive Strategy: Techniques for Analyzing Industries and Competitors*. New York: Free Press.

Porter, M. E. (1990). *Competitive Advantage of Nations: Creating and Sustaining Superior Performance*. New York: Free Press.

Rivest, R. L., Shamir, A., & Adleman, L. (1978). A method for obtaining digital signatures and public-key cryptosystems. *J Commun. ACM, 21*(2), 120-126. doi: 10.1145/359340.359342.

Schlagberger, E. M., Bornmann, L., & Bauer, J. (2016). At what institutions did Nobel laureates do their prize-winning work? An analysis of biographical information on Nobel laureates from 1994 to 2014. *Scientometrics, 109*(2), 723–767.

Sivertsen, G. (2017). Problems and considerations in the design of bibliometric indicators for national performance based research funding systems *Proceedings of the Science, Technology, & Innovation Indicators Conference "Open indicators: innovation, participation and actor-based STI indicators*. Paris, France.

Thor, A., Bornmann, L., Marx, W., & Mutz, R. (2018). Identifying single influential publications in a research field: New analysis opportunities of the CRExplorer. *Scientometrics, 116*(1), 591–608.

Thor, A., Marx, W., Leydesdorff, L., & Bornmann, L. (2016). Introducing CitedReferencesExplorer (CRExplorer): A program for Reference Publication Year Spectroscopy with Cited References Standardization. *Journal of Informetrics, 10*(2), 503-515.

van Noorden, R., Maher, B., & Nuzzo, R. (2014). The Top 100 Papers. *Nature, 514*(7524), 550-553.

Van Rijsbergen, C. J. (1979). *Information Retrieval. Unpublished PhD thesis*. Glasgow, UK: Department of Computing Science, University of Glasgow.

Wang, J. (2013). Citation time window choice for research impact evaluation. *Scientometrics, 94*(3), 851-872. doi: 10.1007/s11192-012-0775-9.

Wang, J. (2014). Unpacking the Matthew effect in citations. *Journal of Informetrics, 8*(2), 329-339. doi: http://dx.doi.org/10.1016/j.joi.2014.01.006.

Wang, J., Veugelers, R., & Stephan, P. E. (2017). Bias Against Novelty in Science: A Cautionary Tale for Users of Bibliometric Indicators. *Research Policy, 46*(8), 1416-1436.

Weiser, M. (1984). Program slicing. *IEEE Transactions on Software Engineering, SE-10*(4), 439-449.

White, H. D., & Griffith, B. C. (1981). Author Cocitation - a Literature Measure of Intellectual Structure. *Journal of the American Society for Information Science, 32*(3), 163-171. doi: DOI 10.1002/asi.4630320302.

Winnink, J. J., Tijssen, R. J. W., & van Raan, A. F. J. (2018). Searching for new breakthroughs in science: How effective are computerised detection algorithms? *Technological Forecasting and Social Change*. doi: https://doi.org/10.1016/j.techfore.2018.05.018.